\documentclass[12pt]{article}
\usepackage[margin=1.1in]{geometry}
\usepackage{subfigure}
\usepackage{amsmath}
\usepackage{amsthm} 
\allowdisplaybreaks
\usepackage{color}
\usepackage{footnote}
\usepackage{algorithm}
\usepackage{algpseudocode}
\usepackage{algorithmicx}
\usepackage{multirow} 
\usepackage{mdwlist}
\usepackage{booktabs}
\usepackage{graphicx}
\usepackage{amssymb}
\usepackage{amsbsy}
\usepackage{array}
\usepackage{longtable}
\usepackage{epstopdf}
\usepackage{pbox}
\usepackage{breqn}
\usepackage{mathrsfs}
\usepackage{multicol}
\usepackage{supertabular}
\usepackage{enumerate}
\usepackage[colorinlistoftodos]{todonotes}
\usepackage{bbm}
\usepackage{bbold}
\usepackage{url}
\usepackage[utf8x]{inputenc}
\usepackage[bookmarks]{hyperref}
\usepackage{color}
\usepackage[normalem]{ulem}
\usepackage[justification=centering]{caption}
\newtheorem{theorem}{Theorem}
\newtheorem{definition}{Definition}
\newtheorem{lemma}{Lemma}
\newtheorem{proposition}{Proposition}

\newtheorem{remark}{Remark}
\usepackage{amsfonts}
\usepackage{stackrel}
\renewcommand{\)}{\right)}
\newcommand{\deleq}{\stackrel{\Delta}{=}}
\renewcommand{\(}{\left(}
\usepackage[title]{appendix}
\DeclareMathOperator{\argmax}{arg\,max}
\newcommand*{\QEDA}{\hfill\ensuremath{\square}}
\begin{document}
\title{\LARGE \bf Signaling Game-based Misbehavior Inspection in \\V2I-enabled Highway Operations }

\author{Manxi Wu, Li Jin, Saurabh Amin, and Patrick Jaillet
\thanks{The authors thank four anonymous reviewers for helpful comments. We are grateful to Prof. Demos Teneketzis and Dr. Jonathan Petit for the valuable suggestions and discussions. This work was supported in part by the Singapore National Research Foundation through the Singapore MIT Alliance for Research and Technology (SMART) Center for Future Mobility (FM), and US National Science Foundation (NSF) grants.}
\thanks{M. Wu is with the Institute for Data, Systems, and Society, S. Amin is with the Department of Civil and Environmental Engineering, P. Jaillet is with the Department of Electrical Engineering and Computer Science, Massachusetts Institute of Technology (MIT), Cambridge, MA, and L. Jin is with the Department of Civil and Urban Engineering, New York University (NYU), Brooklyn, NY, USA, {\tt\small \{manxiwu,jnl, amins,jaillet\}@mit.edu}}%
}
\date{}

\maketitle
\newcommand{\U}{U}
\newcommand{\h}{h}
\renewcommand{\L}{\mathrm{L}}
\newcommand{\I}{\mathrm{I}}
\newcommand{\N}{\mathrm{N}}
\renewcommand{\l}{l}
\renewcommand{\H}{\mathrm{H}}
\newcommand{\qt}{q^\t}
\newcommand{\qth}{q^\t_h}
\newcommand{\qtl}{q^\t_l}
\newcommand{\ph}{p^\t_h}
\newcommand{\pl}{p^\t_l}
\newcommand{\Fh}{F_h}
\newcommand{\Fl}{F_l}
\renewcommand{\d}{d}
\newcommand{\ut}{u^t}
\newcommand{\udH}{\ud_\H}
\newcommand{\udL}{\ud_\L}
\newcommand{\A}{A}
\newcommand{\B}{B}
\newcommand{\Bone}{\B_1}
\newcommand{\Btwo}{\B_2}
\newcommand{\Bthree}{\B_3}
\newcommand{\lamb}{\lambda}
\newcommand{\cl}{c_{\L}^{\theta}}
\newcommand{\ch}{c_{\H}^{\theta}}
\newcommand{\siglbar}{\widehat{\sig}_{\l}^\t}
\newcommand{\muh}{\mu_\H}
\newcommand{\dc}{\Delta c^{\theta}}
\newcommand{\mul}{\mu_\L}
\newcommand{\dcinv}{\(\dc\)^{-1}}
\newcommand{\plbar}{\bar{p}^\t_\l}
\newcommand{\EsigwelH}{\mathbb{E}_{\sigwe}[U^\t_l(\H)]}
\newcommand{\EsigwelL}{\mathbb{E}_{\sigwe}[U^\t_l(\L)]}
\newcommand{\game}{\Gamma}
\newcommand{\sigh}{\sigma^t_h}
\newcommand{\sigl}{\sigma^t_l}
\newcommand{\sigd}{\sigma^d}
\newcommand{\sig}{\sigma}
\newcommand{\sigt}{\sigma^t}
\newcommand{\sighwe}{\sigma^{t*}_{h}}
\newcommand{\siglwe}{\sigma^{t*}_{l}}
\newcommand{\sigdwe}{\sigma^{d*}}
\newcommand{\sigwe}{\sigma^{*}}
\newcommand{\sigtwe}{\sigma^{t*}}
\newcommand{\chsigt}{c_\H(\sigt)}
\newcommand{\clsigt}{c_\L(\sigt)}
\newcommand{\chbar}{\bar{c}_\H}
\newcommand{\clbar}{\bar{c}_\L}
\newcommand{\chsigtwe}{c_\H^\theta(\sigtwe)}
\newcommand{\clsigtwe}{c_\L^\theta(\sigtwe)}
\newcommand{\uth}{u^t_h}
\newcommand{\utl}{u^t_l}
\newcommand{\Esig}{\mathbb{E}_{\sig}}
\newcommand{\Uh}{U^t_h}
\newcommand{\Esigdh}{\mathbb{E}_{\sig}}
\newcommand{\Esigdl}{\mathbb{E}_{\sig}}
\newcommand{\Ul}{U^t_l}
\newcommand{\bH}{\beta(\H)}
\newcommand{\bHh}{\beta(\h|\H)}
\newcommand{\bHl}{\beta(\l|\H)}
\newcommand{\bL}{\beta(\L)}
\newcommand{\bLh}{\beta(\h|\L)}
\newcommand{\bLl}{\beta(\l|\L)}
\newcommand{\ud}{u^d}
\renewcommand{\b}{\beta}
\newcommand{\sigdh}{\sigd_\H}
\newcommand{\sigdl}{\sigd_\L}
\newcommand{\pd}{p^\d}
\newcommand{\bHwe}{\beta^{*}(\H)}
\newcommand{\bweHh}{\beta^{*}(\h|\H)}
\newcommand{\bweHl}{\beta^{*}(\l|\H)}
\newcommand{\bLwe}{\beta^{*}(\L)}
\newcommand{\bweLh}{\beta^{*}(\h|\L)}
\newcommand{\bweLl}{\beta^{*}(\l|\L)}
\newcommand{\bwe}{\beta^{*}}
\newcommand{\sigdhwe}{\sig^{d*}_\H}
\newcommand{\sigdlwe}{\sig^{d*}_\L}

\newcommand{\UdH}{U_\H^d}
\newcommand{\UdL}{U_\L^d}
\newcommand{\s}{s}
\newcommand{\lel}{\lambda_{\l}}
\newcommand{\les}{\lambda_{\s}}
\renewcommand{\a}{\mathrm{a}}
\newcommand{\uL}{u(\L)}
\newcommand{\uS}{u(\S)}
\newcommand{\csp}{c_{\s}(\L)}
\newcommand{\cs}{c_{\s}(\S)}
\newcommand{\pro}{\mathrm{Pr}}
\newcommand{\qs}{q_{\s}}
\newcommand{\ql}{q_{\l}}
\newcommand{\Qs}{Q_{\s}}
\newcommand{\Ql}{Q_{\l}}
\newcommand{\F}{F}
\newcommand{\m}{m}
\newcommand{\cL}{c(\L)}
\newcommand{\cS}{c(\S)}
\newcommand{\deltac}{\Delta_c}
\newcommand{\Qd}{Q_{d}}

\newcommand{\D}{D}
\newcommand{\ul}{V_{\l}}
\newcommand{\us}{V_{\s}}
\newcommand{\T}{T}
\renewcommand{\t}{t}
\newcommand{\sigswe}{\sigma_{\s}^{*}}

\begin{abstract}
Vehicle-to-Infrastructure (V2I) communications are increasingly supporting highway operations such as electronic toll collection, carpooling, and vehicle platooning. In this paper we study the incentives of strategic misbehavior by individual vehicles who can exploit the security vulnerabilities in V2I communications and negatively impact the highway operations. We consider a V2I-enabled highway segment facing two classes of vehicles (agent populations), each with an authorized access to one server (subset of lanes). Vehicles are strategic in that they can misreport their class (type) to the system operator and get an unauthorized access to the server dedicated to the other class. This misbehavior causes additional congestion externality on the compliant vehicles, and thus, needs to be deterred. We focus on an environment where the operator is able to inspect the vehicles for misbehavior. The inspection is costly and successful detection incurs a fine on the misbehaving vehicle. We formulate a signaling game to study the strategic interaction between the vehicle classes and the operator. Our equilibrium analysis provides conditions on the cost parameters that govern the vehicles' incentive to misbehave or not. We also determine the operator's equilibrium inspection strategy.
\end{abstract}

\section{Introduction}
Vehicle-to-Infrastructure (V2I) and Vehicle-to-Vehicle (V2V) communications are commonly regarded as integral features of smart highway systems \cite{varaiya93,papadimitratos09}. With the projected growth of V2I and V2V capabilities, it is expected that they will support important operations such as safety-preserving maneuvers (overtaking), lane management, intersection control, and also enable traffic management with connected/autonomous vehicles \cite{kurzhanskiy15, bergenhem12}, \cite{jin18hscc}. These applications typically require the presence of road-side units (RSUs) that are capable of receiving messages from individual vehicles (i.e., their on-board units (OBUs)), authenticating these messages, and providing relevant information to the neighboring vehicles and/or actuators (e.g., traffic signals). This message exchange is typically supported by the Dedicated Short Range Communications (DSRC) technology, which enables the RSU to gather information such as vehicle identifier, vehicle class, and safety-related data. In recent years, numerous security concerns have been identified in this context \cite{petit15}, \cite{lawson08}, \cite{hubaux04}. Prior research in security of vehicular communications has focused on the identification of vulnerabilities and design of defense solutions to prevent, detect, and respond to various security threats \cite{alpcan2010network}. However, an aspect that has received relatively little attention is the modeling of strategic misbehavior by vehicles that can negatively impact highway operations. 

\begin{figure}[H]
\centering
\includegraphics[width=0.6\textwidth]{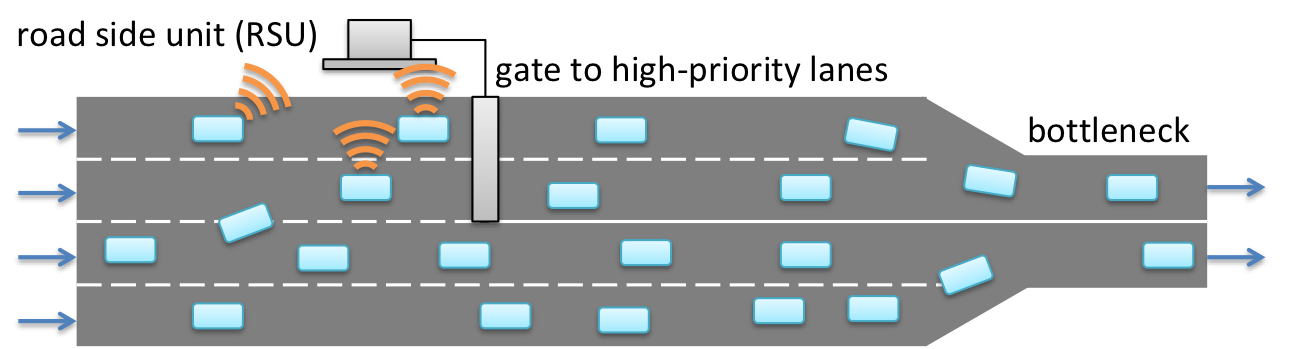}
\caption{A V2I-based highway operations.}
\label{toll_lanes}
\end{figure}

In this paper, we focus on a generic setting of lane management operation enabled by V2I communications, and develop a model of strategic misbehavior using a signaling game formulation. To motivate the setting, let us consider a highway segment with a downstream bottleneck; see Fig.\ref{toll_lanes}. The highway is equipped with a RSU and all incoming vehicles have OBUs. The highway section has two classes of lanes: the high-priority lanes are meant to serve the travelers with preferential access to the system, and the low-priority (or general purpose) lanes are meant to serve all other travelers. We consider the two sets of lanes as parallel servers. The RSU receives and authenticates the messages from the incoming vehicles. A vehicle is provided access to a server if the RSU is able to authenticate its message and adjudge it to be a vehicle belonging to that class. The travel cost incurred in accessing each server increases with the aggregate number of vehicles routed through that server due to the congestion effects. 
We assume that the two classes are pre-established using well-known economic principles.
\vspace{-0.25cm}
The main feature that we consider in the abovementioned setting is the ability of travelers to manipulate the communication between their vehicles' OBU and the RSU, and obtain access to the server that does not correspond to their vehicle class. Such an attack can be realized if the vehicles (henceforth referred to as ``agents'') can misreport their identity information to the RSU. We consider this type of attack as an instance of \emph{strategic misbehavior}, which needs to be deterred by the system operator because it imposes additional congestion externalities on the other travelers. Misbehavior can be deterred
using technological means such as inspecting the messages for their integrity, checking the vehicle's identity using video cameras, number plate recognition, or manual inspection. On successful detection, a suitable fine can be imposed by the operator. It is important to include the following features in modeling misbehavior: (i) The operator has incomplete information about the priority class of incoming vehicles, in that the true class of vehicle can be known only after inspection (which is costly); (ii) Each misbehaving agent incurs a technological cost for manipulating its message, and is subject to a fine if inspected. These features naturally lead us to pose our model as a signaling game \cite{spence1974market}.

In our game, the agents are non-atomic and each agent has private information about its type; i.e., each agent privately knows whether she is a high- or low-priority agent. The operator has the technological capability of message collection (via the RSU), inspection, and fine collection. We say that an agent misbehaves if she sends a signal that is different from her true type and obtains access to the lane (server) that does not correspond to her true type. 
All agents are subject to inspection by the operator. A misbehaving agent incurs a non-negative cost, and if detected, is charged a non-negative fine. 
	
	The equilibrium concept that we use is the Perfect Bayesian Equilibrium (PBE) \cite{fudenberg1991perfect}. In the PBE, (a) the players satisfy sequential rationality, and (b) the operator's belief is consistent with the prior distribution of agent types and the agents' strategy. The specific features that distinguish our game from the classical models of signaling games are: non-atomic agent populations, and congestion externality imposed by an agent on other agents sending the same signal (i.e. accessing the same server). Under mild assumptions on the cost functions of both servers, we provide a complete characterization of the PBE for our signaling game. 
	
	In particular, we show that in equilibrium (i) a high priority agent does not have the incentive to misbehave for gaining access to the low-priority server; (ii) not every low-priority agent misbehaves. Moreover, we distinguish two regimes based on how the technological cost of misbehavior compares with the maximum gain from misbehavior (evaluated as the difference in travel costs of two servers when no agent misbehaves). In the first regime, the misbehavior is completely deterred as the technological cost of misbehavior is high and the operator does not need to inspect any agent. In the second regime, the low-priority agents misbehave with a positive probability. The operator's inspection strategy in the second regime can be further distinguished by sub-regimes that correspond to zero, partial, and complete inspection.  
	
Our equilibrium analysis can be used to study the comparative statics with respect to practically relevant parameters such as the fraction of each priority class, the inspection cost, and the fine. Firstly, for given cost of misbehavior and cost of inspection, the equilibrium misbehavior rate (and the hence the rate of inspection) decreases as the fraction of high-priority agents increases. Secondly, fine can be effective for decreasing misbehavior rate even when the inspection cost is high (relative to the cost of misbehavior), but cannot achieve complete deterrence. Thirdly, if the fine is sufficiently high, then there is no need to inspect all agents in equilibrium. These insights are relevant for the design and deployment of inspection technologies to achieve higher security levels of V2I-enabled highway operations. Finally, we illustrate these effects in the setting of Electronic Toll Collection (ETC), where the servers are modeled as M/M/1 queueing systems, and the fraction of high-priority travelers (and the toll that they need to pay for priority access) is exogenously known.

\section{Modeling Misbehavior} \label{sec_physical}

In this section, we consider a model of lane management operations on a highway section equipped with vehicle-to-infrastructure (V2I) communications capability, and propose a model of strategic misbehavior for this setting.

Suppose that the highway system faces a fixed traffic demand comprised of two types of agent populations: a high priority type, denoted $\h$, and a low priority type, denoted $\l$. 
The fraction of type $\h$ agents is $\theta \in (0,1)$, and the fraction of type $\l$ agents is $1-\theta$. 
Throughout this article, we assume that $\theta$ is exogenous and independent of potential misbehavior; see Remark \ref{r1}.
There are two sets of lanes on the highway, $\H$ and $\L$, which we model as two parallel servers. 
In the absence of any misbehavior, server $\H$ is only assessed by $\h$ agents, and server $\L$ by the type $\l$ agents.
\begin{remark}\label{r1}
 Admittedly, the assumption of fixed fractions $\theta$ precludes us from considering situations where the agent populations would choose their priority type (routing behavior) in anticipation of the potential misbehavior that they may face. However, it allows us to identify the effect of any given $\theta$ on the misbehavior rate. 
 \end{remark}

To reduce his/her travel cost, an agent may manipulate his reported signal, and choose to take the server that is not meant for his type. 
We use $\sigl$ (resp. $\sigh$) to denote the fraction of $\l$ (resp. $\h$) agents that misbehave.
We denote the misbehavior strategy as $\sigt=(\sigh,\sigl)$. 
Since the aggregate demand of agents using a server is determined by the relative size of agent populations ($\theta$), and misbehavior strategy profile ($\sigt$), we will use the notations $\ch(\sigt)$ (resp. $\cl(\sigt)$) to denote the cost of server $\H$ (resp. $\L$).

We make the following assumptions on the cost functions:
\begin{enumerate}
\item[(A1)] $\ch(0,0)<\infty$, $\cl(0,0)<\infty$, $\ch(0,0)<\cl(0,0)$, and $\ch(0, 1) > \cl(0, 1)$
\item[(A2)] $\ch(\sigh, \sigl)$ (resp. $\cl(\sigh, \sigl)$) decreases in $\sigh$ (resp. $\sigl$), and increases in $\sigl$ (resp. $\sigh$) 
\item[(A3)] $\ch(\sigh, \sigl)$ increases in $\theta$. $\cl(\sigh, \sigl)$ decreases in $\theta$.
\end{enumerate}
(A1) reflects that server $\H$ has higher priority than server $\L$, and ensures that both servers face stable queues. (A2) and (A3) captures the congestion nature of the highway system. 

\section{Signaling game for misbehavior inspection}\label{sec_misbehavior}
We now model the strategic interaction between the agent populations (travelers $\t$) that are prone to misbehavior and the system operator (defender $\d$) who decides to inspect them based on the received messages. We consider that the agents are capable of compromising the integrity of messages sent to the RSU in order to obtain access to the server that does not correspond to their true type. Recall that
the operator cannot know an agent's true type unless she inspects the agent. This information asymmetry between the agents and the operator naturally leads to a signaling game formulation. 

In the game, each agent type is modeled as a population of non-atomic players. The cost of misbehavior is non-negative for each type $\h$ agent, denoted $\ph \in \mathbb{R}_{\geq 0}$, and strictly positive for each type $\l$ agent, denoted $\pl \in \mathbb{R}_{> 0}$.\footnote{We make this technical assumption to avoid triviality in equilibrium analysis. It is consistent with our setting of differentiated priority system.}

As mentioned before, the operator does not know the type of each agent, but can observe the agent's signal, i.e. the server taken by the agent. The signal space is the set of servers $\{\H,\L\}$. We say that a type $\l$ (resp. type $\h$) agent misbehaves if she chooses the server $\H$ (resp. $\L$). The operator forms a belief of the true type given the observed signal. We denote $\bH \deleq \(\bHh, \bHl\)$ (resp. $\bL \deleq \(\bLh, \bLl\)$) as the operator's belief given the signal $\H$ (resp. $\L$), where $\bHh$ and $\bHl$ (resp. $\bLh$ and $\bLl$) are the posterior probabilities that an agent on the server $\H$ (resp. $\L$) is in fact a type $\h$ and $\l$ agent, respectively. Based on the signal and the belief, the operator chooses to inspect an agent ($\I$), or not to inspect ($\N$). The inspection incurs a positive cost on the operator, denoted $\pd \in \mathbb{R}_{>0}$. We denote $\sigdh$ (resp. $\sigdl$) as the probability of inspecting an agent on the server $\H$ (resp. $\L$). Then, the operator's inspection strategy is $\sigd \deleq \(\sigdh, \sigdl\)$, and the strategy profile is $\sig \deleq \(\sigt, \sigd\)$. Furthermore, for simplicity, we assume that if an agent misbehaves, and if he is inspected, then the misbehavior is detected with probability 1.\footnote{If the probability of
detection is smaller than 1, then the effective inspection rate is simply the total inspection rate scaled by the detection probability. Our analysis approach can be straightforwardly extended to this case.} A fine $\Fh \in \mathbb{R}_{\geq 0}$ (resp. $\Fl \in \mathbb{R}_{\geq 0}$) is charged to the type $\h$ (resp. $\l$) agent if his misbehavior is detected. 

We are now ready to discuss the utility functions of the agents and the operator. The utility of each agent is the summation of three parts: (i) $-\ch(\sigt)$ (resp. $-\cl(\sigt)$), which is the travel cost if the agent chooses the server $\H$ (resp. $\L$); (ii) $-\ph$ (resp. $-\pl$), which is the technology cost of misbehavior for a type $\h$ (resp. $\l$) agent; (iii) $-\Fh$ ($-\Fl$), which is the fine if the misbehavior is detected upon inspection of a type $\h$ (resp. $\l$) agent. The utility of the operator is the summation of two parts: (i) $-\pd$, which is the inspection cost; (ii) $\Fh$ (resp. $\Fl$), which is collected fine when the misbehavior of a type $\h$ (resp. $\l$) agent is detected.

\begin{figure}[htp]
\centering
\includegraphics[width=0.7\textwidth]{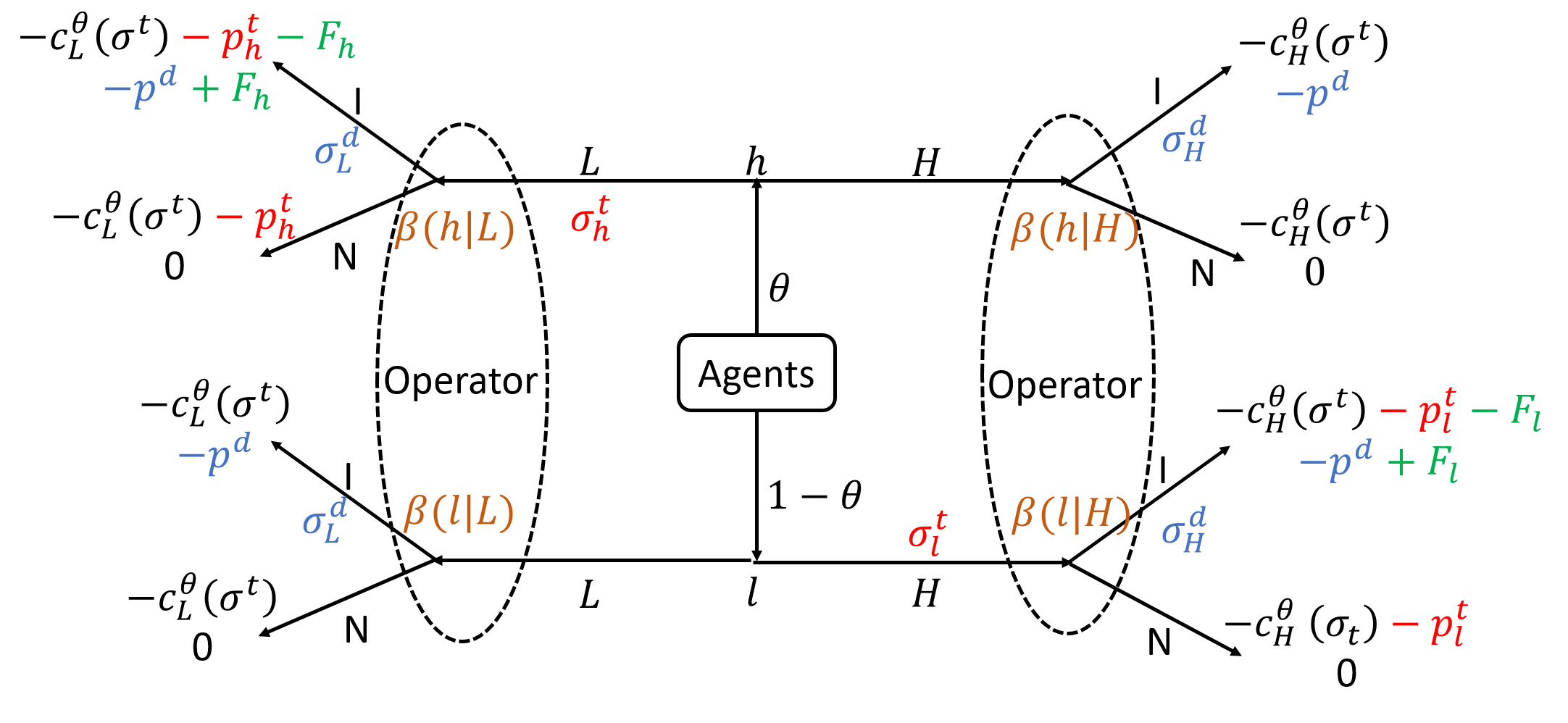}
\caption{Game Tree with the agent's utility (top) and the operator's utility (bottom) indicated at each leaf node.}
\label{game_diagram}
\end{figure}

The game is played in the following steps; see Fig. \ref{game_diagram}. First, the type of each agent is chosen by the fictitious player ``Nature'' according to the probability distribution $\pro(\h)=\theta$ and $\pro(\l)=1-\theta$. Then the agents send the signal (choose the server) according to strategy $\sigt$ based on their type. Next, the operator observes the signal, and the belief $\b$ is updated based on the observed signal. The operator then chooses to inspect or not according to $\sigd$. All the game parameters are common knowledge, except that each agent privately knows his type. 

Given strategy profile $\sig$, we denote the expected utility of type $\h$ agents choosing the server $\H$ (resp. $\L$) as $\Esig[\Uh(\H)]$ (resp. $\Esig[\Uh(\L)]$). The expected utilities for type $\l$ agents are similarly denoted as $\Esig[\Ul(\H)]$ and $\Esig[\Uh(\H)]$, respectively. The expected utilities of agents can be written as follows:

\begin{footnotesize}
\begin{subequations}\label{expected_cost_agent}
\begin{align}
\Esig[\Uh(\H)]&=-\ch(\sigt),\text{   } \Esig[\Uh(\L)]=-\cl(\sigt)-\ph-\Fh \sigdl, \label{UhL}\\
\Esig[\Ul(\H)]&=-\ch(\sigt)-\pl-\Fl \sigdh, \text{   } \Esig[\Ul(\L)]=-\cl(\sigt). \label{UlL}
\end{align}
\end{subequations}
\end{footnotesize}Given any strategy profile $\sig$ and any belief $\b$, the expected utility of the operator when observing signal $\H$ (resp. $\L$) is denoted as $\Esigdh[\UdH|\beta]$ (resp. $\Esigdl[\UdL|\beta]$), which can be written as follows:
\begin{footnotesize}
\begin{subequations}\label{expected_cost_operator}
\begin{align}
\Esigdh[\UdH|\beta] &= \(-\pd+ \Fl  \bHl \) \sigdh, \label{cost_H}\\
\Esigdl[\UdL|\beta]&= \(-\pd + \Fh \bLh\) \sigdl.\label{cost_L}
\end{align}
\end{subequations}
\end{footnotesize}Note that in all the cost functions \eqref{UhL}-\eqref{UlL} and \eqref{cost_H}-\eqref{cost_L}, the travel cost is the expected travel time multiplied with the value of time, and hence can be treated as a monetary cost, similar to the fine and misbehavior cost.

The equilibrium concept in this game is the perfect Bayesian equilibrium (PBE), see \cite{fudenberg1991perfect}:
\begin{definition}
A pair $(\sigwe, \bwe)$ of strategy profile $\sigwe$ and belief assessment $\bwe$ is a PBE if it satisfies both sequential rationality and consistency:
\begin{itemize}
\item \emph{Sequential rationality}: (i) The servers that are used by each type of agents incur the highest expected utility: 
\begin{footnotesize}
\begin{subequations}\label{rational_traveler}
\begin{align}
\sighwe >0 &\quad \Rightarrow \quad \mathbb{E}_{\sigwe}[\Uh(\L)] \geq \mathbb{E}_{\sigwe}[\Uh(\H)], \label{travel_high_deviate}\\
\sighwe <1 &\quad \Rightarrow \quad \mathbb{E}_{\sigwe}[\Uh(\L)] \leq \mathbb{E}_{\sigwe}[\Uh(\H)], \label{travel_high}\\
\siglwe >0 &\quad  \Rightarrow \quad \mathbb{E}_{\sigwe}[\Ul(\H)] \geq \mathbb{E}_{\sigwe}[\Ul(\L)], \label{travel_low_deviate}\\
\siglwe <1 &\quad  \Rightarrow \quad \mathbb{E}_{\sigwe}[\Ul(\H)] \leq \mathbb{E}_{\sigwe}[\Ul(\L)]. \label{travel_low}
\end{align} 
\end{subequations}
\end{footnotesize}
(ii) The operator maximizes expected utility:
\begin{footnotesize}
\begin{equation}\label{eq_operator_rational}
\begin{split}
\sigdhwe &= \argmax_{\sigdh \in [0, 1]} \Esigdh[\UdH|\bwe], \text{  } \sigdlwe=\argmax_{\sigdl \in [0, 1]} \Esigdl[\UdL|\bwe]
\end{split}
\end{equation}
\end{footnotesize}
\item \emph{Consistency}: $\bwe$ is updated according to the agent's strategy $\sigwe$ using the Bayes' rule:
\begin{footnotesize}
\begin{subequations}\label{bayes}
\begin{align}
\bweHh &= \frac{\theta  \(1-\sighwe\)}{\theta  \(1-\sighwe\)+ \(1-\theta\) \siglwe},  \\
\bweLh &= \frac{\theta  \sighwe}{\theta  \sighwe+\(1-\theta\) \(1-\siglwe\)},
\end{align} 
\end{subequations}
\end{footnotesize}
and $\bweHl=1-\bweHh$, $\bweLl=1-\bweLh$. 
\end{itemize}
\end{definition}

We offer two remarks on PBE. First, with regard to the sequential rationality of agents, the rationality constraints \eqref{travel_high_deviate} and \eqref{travel_low_deviate} ensure that if the agents of a given type misbehave with positive probability in equilibrium, then the expected utility in choosing to access the other server is no less than the expected utility in choosing to access the server corresponding to their own type. On the other hand, constraints \eqref{travel_high} and \eqref{travel_low} ensure that if agents use the server for their true type with positive probability in equilibrium, then the utility of choosing the other server is not strictly higher.

Second, the consistency of beliefs requires that the operator's updated belief of each type given the received signal is consistent with the prior distribution and the agents' strategy in accordance with the Bayes' rule. The operator then chooses the optimal inspection rate based on her belief.

\section{Equilibrium Characterization}\label{sec_equilibrium}
In this section, we characterize the PBE of the signaling game. In Sec. \ref{properties}, we show three properties of PBE that are crucial for equilibrium analysis. In Sec. \ref{regimes}, we focus on analyzing the equilibrium regimes, where the properties of PBE are qualitatively distinct. 

\subsection{General properties of PBE}\label{properties}
From (A1), we know that the costs of both servers are finite when no agent misbehaves. The following lemma guarantees stability in PBE, i.e. the costs of both servers are also finite in equilibrium. 
\begin{lemma}\label{finite_equilibrium}
In any PBE, $\ch(\sigtwe)<\infty$ and $\cl(\sigtwe)<\infty$. 
\end{lemma}
\begin{proof}
We prove by contradiction. If $\ch(\sigtwe)=\infty$, the aggregate amount of agents taking server $\H$ in equilibrium must be higher than that without misbehavior. Hence, the amount of agents on server $\L$ is lower than that without misbehavior, which ensures $\cl(\sigtwe)<\infty$. Given any operator's strategy $\sigd \in [0,1]$, from \eqref{expected_cost_agent}, we must have $\mathbb{E}_{\sigwe}[\Uh(\H)] < \mathbb{E}_{\sigwe}[\Uh(\L)]$ and $\mathbb{E}_{\sigwe}[\Ul(\H)]< \mathbb{E}_{\sigwe}[\Ul(\L)]$. From \eqref{travel_high} and \eqref{travel_low_deviate}, we have $\sighwe=1$ and $\siglwe=0$, i.e. no agents use server $\H$ in equilibrium. This contradicts the claim that $\ch(\sigtwe)=\infty$. Analogously, we argue that $\cl(\sigtwe)<\infty$.  
\end{proof}
The next proposition shows that type $\h$ agents do not misbehave in equilibrium. Consequently, the operator does not inspect agents that choose to access the server $\L$. 
\begin{proposition}\label{h_non_deviate}
In any PBE, $(\sigwe, \bwe)$ satisfies:\vspace{-0.1cm}
\[\sighwe=0, \quad \sigdlwe=0, \quad \bweLl=1, \quad \bweLh=0.\]
\end{proposition}
\begin{proof}
We first prove $\sighwe=0$ by contradiction. Assume that there exists a PBE such that $\sighwe>0$, i.e. there exists a fraction of type $\h$ agents using server $\L$.  From \eqref{UhL} and \eqref{travel_high_deviate}, we must have $-\clsigtwe-\ph \stackrel{\eqref{UhL}}{\geq} \mathbb{E}_{\sigwe}[\Uh(\L)] \stackrel{\eqref{travel_high_deviate}}{\geq} \mathbb{E}_{\sigwe}[\Uh(\H)]\stackrel{\eqref{UhL}}{=} -\chsigtwe$. Thus, $\clsigtwe \leq \chsigtwe -\ph$. Since $\ph \geq 0$, $\clsigtwe \leq \chsigtwe$. From \eqref{UlL}, we have $\mathbb{E}_{\sigwe}[\Ul(\L)] \stackrel{\eqref{UlL}}{=} -\clsigtwe \geq -\chsigtwe \stackrel{\eqref{UlL}}{>} \mathbb{E}_{\sigwe}[\Ul(\H)]$. Hence, from \eqref{travel_low_deviate}, we must have $\siglwe=0$, i.e. no agents of type $\l$ take server $\H$. Additionally, since $\clsigt$ is increasing in $\sigh$ and decreasing in $\sigl$, when $\sighwe>0$ and $\siglwe=0$, we have $\clsigtwe > \cl(0,0)$. Analogously, $\chsigt$ is increasing in $\sigl$ and decreasing in $\sigh$, and thus $\chsigtwe < \ch(0,0)$.  Consequently, we derive $\ch(0,0) > \chsigtwe \geq \clsigtwe > \cl(0,0)$, which contradicts (A1). Therefore, $\sighwe=0$. 

Next, from \eqref{bayes}, we can check that the belief updated by Bayes' rule satisfies $\bweLl=1$ and $\bweLh=0$. 

Finally, since $\bwe(\l|\L)=1$ implies that only type $\l$ agents take server $\L$. From \eqref{cost_L}, the action $\I$ is strictly dominated by the action $\N$. Hence, $\sigdlwe=0$.  \end{proof}

In addition, the next proposition ensures that not all type $\l$ agents misbehave in equilibrium. 
\begin{proposition}\label{no_pool}
In any PBE, $\siglwe<1$. 
\end{proposition}
\begin{proof}
Again we prove this claim by contradiction. Assume that $\siglwe=1$, i.e. all the agents of type $\l$ uses server $\H$. From Proposition \ref{h_non_deviate}, agents of type $\h$ do not use server $\L$ in equilibrium. Therefore, in PBE, no agents use server $\L$. From (A1), we know that $\mathbb{E}_{\sigwe}[\Ul(\L)]=-\cl(0,1)>-\ch(0,1) \geq \mathbb{E}_{\sigwe}[\Ul(\H)]$, which contradicts the equilibrium condition in \eqref{travel_low_deviate}. Hence, $\siglwe<1$.   \end{proof}

From Propositions \ref{h_non_deviate} -- \ref{no_pool}, we can conclude that both servers are used in equilibrium. This implies that ``pooling'' equilibrium, in which agents of different types send identical signals, does not exist in our game. Consequently, we can drop $\sighwe$, $\bwe(\cdot|\L)$ and $\sigdlwe$ in our further analysis. For simplicity, we abuse the notation and use $\ch(\sigl)$ (resp. $\cl(\sigl)$) to denote the cost of server $\H$ (resp. $\L$) when type $\l$ agents' strategy is $\sigl$, and $\sigh=0$. Additionally, we define $\dc(\sigl)$ as the cost difference between $\L$ and $\H$ when the strategy of type $\l$ agents is $\sigl$: 
\begin{align*}
\dc(\sigl) \deleq \cl(\sigl)-\ch(\sigl).
\end{align*} 
The function $\dc(\sigl)$ evaluates the incentive of a type $\l$ agent to misbehave given that the fraction of misbehaving type $\l$ agents is $\sigl$. From Assumptions (A2)-(A3), we know that this incentive strictly decreases in type $\l$'s misbehavior rate $\sigl$ and the fraction of type $\h$ agents $\theta$. We can thus properly define the function $\dcinv$, which is the inverse of $\dc$.

\subsection{Equilibrium regimes}\label{regimes}
We now provide a full characterization of PBE. From Propositions \ref{h_non_deviate}-\ref{no_pool}, we know that in general, there are two possible cases in equilibrium: In the first case, no agent of type $\l$ takes server $\H$, which means misbehavior is completely deterred; and in the other case, a fraction of type $\l$ agent population takes server $\H$. Indeed, we find that there exist two equilibrium regimes, each corresponding to one of the two cases. Furthermore, the second regime (i.e. a positive fraction of type $\l$ agents take server $\H$) can be divided into three subregimes depending on whether the inspection rate of the operator on server $\H$ is zero, positive or one. 

Before presenting the equilibrium structure, we introduce the following misbehavior threshold rate:
\begin{align}\label{siglbar}
\siglbar \deleq \frac{\pd \theta}{(1-\theta) (\Fl-\pd)}.
\end{align}
Note that the threshold $\siglbar$ is dependent on the inspection cost $\pd$, the fine $\Fl$, and the relative class size $\theta$. 

The next lemma provides the best response correspondence $\sigdhwe$ in equilibrium. 
\begin{lemma}\label{BR}
Given any PBE, if $0<\pd \leq (1-\theta) \Fl$, and 
\begin{itemize}
\item[-] If $\siglwe<\siglbar$, then $\bwe(\l|\H)<\pd/\Fl$ and $\sigdhwe=0$
\item[-] If $\siglwe=\siglbar$, then $\bwe(\l|\H)=\pd/\Fl$ and $\sigdhwe\in [0,1]$
\item[-] If $\siglwe>\siglbar$, then $\bwe(\l|\H)>\pd/\Fl$ and $\sigdhwe=1$.
\end{itemize}
Additionally, if $\pd>(1-\theta) \Fl$, then $\sigdhwe=0$.
\end{lemma}
\begin{proof}
First, we can check that if $0 < \pd \leq (1-\theta) \Fl$, then $\siglbar \in [0, 1]$. From \eqref{bayes}, we know that if $\siglwe=\siglbar$, then $\bwe(\l|\H)= \pd/\Fl$. In this case, $-\pd+\Fl \bwe(\l|\H)=0$, and thus any $\sigdhwe \in [0,1]$ maximizes $\Esigdh[\UdH|\bwe]$ in \eqref{cost_H}. Additionally, since $\bwe(\l|\H)$ increases in $\siglwe$, if $\siglwe<\siglbar$, we must have $\bwe(\l|\H)< \pd/\Fl$. Consequently, $-\pd+\Fl \bwe(\l|\H)<0$, and from \eqref{cost_H} and \eqref{eq_operator_rational}, $\sigdhwe=0$. The case for $\siglwe>\siglbar$ can be argued analogously. 

Additionally, we argue for the case in which $\pd>(1-\theta)\Fl$. Following from \eqref{bayes} and the fact that $\siglwe \in [0, 1]$ as well as $\sighwe=0$, we have $\bwe(\l|\H) =1-\frac{\theta}{\theta+(1-\theta) \siglwe} \leq 1-\theta.$ Hence, if $\pd>(1-\theta)\Fl$, then $-\pd + \Fl \bwe(\l|\H) <0$. From \eqref{cost_H} and \eqref{eq_operator_rational}, we know that $\sigdhwe=0$. 
\end{proof}

Lemma \ref{BR} shows that the probability of detecting a misbehavior on server $\H$ is no higher than $(1-\theta)$, which is achieved only when all type $\l$ agents misbehave. Therefore, the maximum expected fine is $(1-\theta)\Fl$. If the inspection cost is higher than the maximum expected fine, then the operator does not inspect any agent. On the other hand, if the inspection cost is lower than the maximum expected fine, then following the belief update rule \eqref{bayes}, $\siglbar$ leads to the belief $\bwe(\l|\H)=\pd/\Fl$, which is the threshold belief such that the operator is indifferent between the action $\I$ and $\N$ in equilibrium. If $\siglwe$ is higher (resp. lower) than $\siglbar$, then the operator inspects the agents taking the server $\H$ with probability one (resp. zero). 

Additionally, note that as the fraction of $\h$ type goes to zero (i.e. $\theta \to 0$), the threshold $\siglbar \to 0$, which implies that the defender will tend to inspect with probability 1. This is intuitive because when the fraction of type $\h$ is small, even if the misbehavior rate of type $\l$ agent is low, the operator still has a high chance of detecting a misbehavior by inspecting agents on the server $\H$. 

We next introduce the equilibrium regimes in terms of the misbehavior cost $\pl$ and the inspection cost $\pd$:\footnote{Due to space limitations, we only discuss generic cases, where the game parameters are in the interior of each regime. }
\begin{enumerate}
\item In regime $\A$,  $\pl$ satisfies $\pl >\dc(0)$.
\item In regime $\B$, $\pl$ satisfies $\pl <\dc(0)$.
There are three subregimes of regime $\B$. \\
Subregime $\Bone$: 

\begin{footnotesize}
\begin{equation}\label{B1}
\begin{split}
&\left\{\(\pl, \pd\) \left\vert 
\begin{array}{l}
 \max \{\dc(\siglbar), 0\} < \pl <  \dc(0), \text{and}\\
 0<\pd<(1-\theta)\Fl
 \end{array}
 \right.\right\} \bigcup \left\{\(\pl, \pd\) \left\vert 
\begin{array}{l}
0 < \pl <  \dc(0), \text{and}\\
\pd> (1-\theta)\Fl
 \end{array}
 \right.\right\}
\end{split}
\end{equation}
\end{footnotesize}Subregime $\Btwo$:
\begin{footnotesize}
\begin{align}\label{B_2_condition}
\noindent\left\{\(\pl, \pd\) \left\vert\begin{array}{l}
 \max \{\dc(\siglbar)-\Fl, 0\} < \pl\\
  <  \max\{\dc(\siglbar), 0\}, \text{and}\\
  0<\pd<(1-\theta)\Fl.
  \end{array}
  \right.\right\}
\end{align}
\end{footnotesize}Subregime $\Bthree$:
\begin{footnotesize}
\begin{align}\label{condition_B_3}
\left\{\(\pl, \pd\) \left\vert 
\begin{array}{l}
 0< \pl < \max \{\dc(\siglbar)-\Fl, 0\},\text{and}\\
 0<\pd<(1-\theta)\Fl.
 \end{array} \right.\right\}
\end{align}
\end{footnotesize}
\end{enumerate}
Note that the boundaries of subregimes are non-linear since the threshold $\siglbar$ is non-linear in the inspection cost $\pd$. The interpretations of regime boundaries will be straightforward after we present the PBE in each regime. Note that the regime A, and subregimes B1 and B2 are non-empty, but subregime $\Bthree$ can be empty if the fine $\Fl$ is sufficiently high. 

We are now ready to fully characterize the PBE.
\begin{theorem}\label{theorem_equilibrium}
PBE is unique in each regime, and can be written as follows:
\begin{enumerate}
\item[(a)] Regime $\A$: 
\begin{footnotesize}
\begin{align}\label{eq_A}
\siglwe=0, \quad \sigdhwe=0, \quad \bwe(\h|\H)=1, \quad \bwe(\l|\H)=0.
\end{align}
\end{footnotesize}
\item[(b)] Regime $\B$:\\
\underline{Subregime $\Bone$}: 
\begin{footnotesize}
\begin{subequations}\label{eq_B_1_general}
\begin{alignat}{2}
&\siglwe=\dcinv(\pl), && \quad \sigdhwe=0, \label{eq_B_1}\\
&\bwe(\h|\H)=\frac{\theta}{\theta+ (1-\theta) \siglwe}, &&\quad \bwe(\l|\H)=\frac{(1-\theta) \siglwe}{\theta+ (1-\theta) \siglwe} \label{b_B_1}
\end{alignat}
\end{subequations}
\end{footnotesize}
\underline{Subregime $\Btwo$}:
\begin{footnotesize}
\begin{subequations}\label{eq_B_2_general}
\begin{alignat}{2}
&\siglwe= \siglbar, &&\quad \sigdhwe=\frac{\dc(\siglbar)-\pl}{\Fl}, \label{eq_B_2}\\
&\bwe(\h|\H)=\frac{\Fl-\pd}{\Fl},&& \quad \bwe(\l|\H)=\frac{\pd}{\Fl} 
\end{alignat}
\end{subequations}
\end{footnotesize}
\underline{Subregime $\Bthree$}: 
\begin{footnotesize}
\begin{subequations}
\begin{alignat}{2}
&\siglwe=\dcinv(\pl+\Fl), \quad && \sigdhwe=1,\label{eq_B_3}\\
&\bwe(\h|\H)=\frac{\theta}{\theta+ (1-\theta) \siglwe}, \quad && \bwe(\l|\H)=\frac{(1-\theta) \siglwe}{\theta+ (1-\theta) \siglwe}\label{b_B_3}
\end{alignat}
\end{subequations}
\end{footnotesize}
\end{enumerate}
\end{theorem}
The proof of this theorem is in Appendix A.

Now we discuss the equilibrium structure in detail. First, we interpret the regime boundaries:
\begin{enumerate}
\item[(i)] Regimes $\A$ and $\B$ are distinguished by the threshold $\dc(0)$, which is the travel cost reduction that a type $\l$ agent can enjoy by misbehaving given that all the other agents are complaint.
\item[(ii)] In regime $\B$, the threshold $(1-\theta) \Fl$ is the maximal expected fine obtained by the operator. We say that the inspection cost is high if $\pd$ is higher than $(1-\theta) \Fl$, and low otherwise. 

Additionally, the threshold $\dc(\siglbar)$ is the gain from misbehavior when the misbehavior rate is $\siglbar$ and the operator does not inspect any agent, and the threshold $\dc(\siglbar)-\Fl$ is the gain when the operator inspects all agents on the server $\H$. We say that the misbehavior cost $\pl$ is \emph{relatively high} compared to $\pd$, if $\pl> \dc(\siglbar)$; \emph{relatively medium} if $\dc(\siglbar)-\Fl<\pl < \dc(\siglbar)$, and \emph{relatively low} if $\pl < \dc(\siglbar)-\Fl$.
\end{enumerate}

We can relate the equilibrium strategy profiles and the conditions determining the regime boundaries:
\begin{basedescript}{\desclabelstyle{\pushlabel}}
\item{ [Regime $\A$]: Misbehavior cost $\pl > \dc(0)$. Misbehavior is fully deterred, and no inspection is needed.}
\item{[Regime $\B$]: Misbehavior cost $\pl<\dc(0)$. Misbehavior occurs with positive probability. }
\end{basedescript}
\begin{itemize}
\item[-] $\Bone$: \emph{Misbehavior cost is relatively high or the inspection cost is high.} The operator does not inspect. The misbehavior rate is such that server $\H$ and $\L$ incur identical utility for type $\l$ agents given that no agent is inspected.
\item[-] $\Btwo$: \emph{Misbehavior cost is relatively medium and the inspection cost is low.} The operator inspects a positive fraction of agents. Misbehavior rate is equal to the threshold $\siglbar$ in \eqref{siglbar}. 
\item[-] $\Bthree$: \emph{Misbehavior cost is relatively low and the inspection cost is low.} The operator inspects all the agents. The misbehavior rate is such that server $\H$ and $\L$ incur identical utility for type $\l$ agents given that all agents on server $\H$ is inspected. Moreover, this rate is higher than the threshold $\siglbar$. 
\end{itemize}



We summarize how PBE changes with the misbehavior and inspection costs in table \ref{comparison}:
\begin{table}[H]
\centering
\begin{tabular}{|c|c|c|c|c|c|}
\hline
\multicolumn{2}{|c|}{} & $\A$ &  $\Bone$ & $\Btwo$ & $\Bthree$\\
\hline
\multirow{2}{*}{$\pl$ increases} & $\siglwe$ & $-$ & $\downarrow$& $-$ &$\downarrow$ \\
\cline{2-6}
&$\sigdhwe$ & $-$ & $-$ & $-$ & $-$\\
\hline
\multirow{2}{*}{$\pd$ increases} & $\siglwe$ & $-$ & $-$ & $\uparrow$  & $-$ \\
\cline{2-6}
&$\sigdhwe$ & $-$ & $-$ & $\downarrow$ & $-$\\
\hline
\end{tabular}
\caption{Qualitative properties of PBE}
\label{comparison}
\end{table}
\vspace{-0.4cm}
The main implications of our equilibrium analysis are as follows: 
\begin{itemize}
\item[-] The misbehavior is completely deterred only when the misbehavior cost is sufficiently high.
\item[-] In subregime $\Btwo$, the belief $\bwe(\H)$ does not depend on $\theta$. The intuition is that in this subregime, both the agents and the operator use mixed strategies in equilibrium, thus, the threshold strategy $\siglbar$ in \eqref{siglbar} increases in $\theta$ to ensure that the belief $\bwe(\l|\H)$ (resp. $\bwe(\h|\H)$) is maintained at the threshold value $\pd/\Fl$ (resp. $1-\pd/\Fl$), which makes the operator indifferent between $\I$ and $\N$. 
\item[-] One can verify the intuitive property that the utility of the type $\l$ (resp. $\h$) agents is non-increasing (resp. non-decreasing) in $\pl$ and non-decreasing (resp. non-increasing) in $\pd$. Similarly, the operator's utility is non-decreasing in $\pl$ and non-increasing in $\pd$. 
\item[-] In general, the misbehavior rate $\siglwe$ is non-increasing in $\pl$, and non-decreasing in $\pd$. The inspection rate $\sigdhwe$ is non-decreasing in $\pl$, and non-increasing in $\pd$.
\end{itemize}

Finally, from (A3), we know that the minimal technology cost needed to deter misbehavior, $\dc(0)$, decreases in $\theta$. Also, $\dc(\siglbar)$ decreases in $\theta$. Therefore, as $\theta$ increases, the sizes of the regime $\A$ and the subregime $\Bone$ increase, and the sizes of the two other subregimes decrease. This implies that the misbehavior rate is lower and the inspection is less needed when more agents are of type $\h$. 

Moreover, as the fine $\Fl$ increases, the size of $\Btwo$ increases, and the size of $\Bthree$ decreases or becomes empty. However, $\Fl$ has no effect on $\A$ and $\Bone$, where inspection is not needed. This observation implies that (i) Fine is effective in reducing the misbehavior rate when the inspection cost is relatively high compared to the misbehavior cost, but cannot fully deter misbehavior, and (ii) If the fine is higher than $\dc(0)$, then the operator will not inspect all agents.

\section{Toll Evasion Example}\label{sec_implication}
We apply our equilibrium results to a specific example of Electronic Toll Collection (ETC) system. In the ETC setting, server $\H$ (resp. $\L$) represents the tolled (resp. toll-free) lanes. Type $\h$ are the agents that are willing to pay the toll, and type $\l$ are the agents that are not willing to pay. The total arrival rate of both types of agents is $\lamb=2400$ veh/hr. The fraction $\theta=0.3$ is the fraction of type $\h$ agents.
Therefore, the arrival rate of type $\h$ (resp. $\l$) agents is $\theta\lambda$ (resp. $(1-\theta)\lambda$). 

The toll is collected electronically, and the access to the tolled lanes is granted to the paying agents after the RSU obtains their reported identifier. 
Such an operation is technologically feasible; see e.g. the European DSRC Toll Collection systems \cite{papadimitratos09}. 

\begin{figure}[H]
    \centering
        \includegraphics[width=0.5\textwidth]{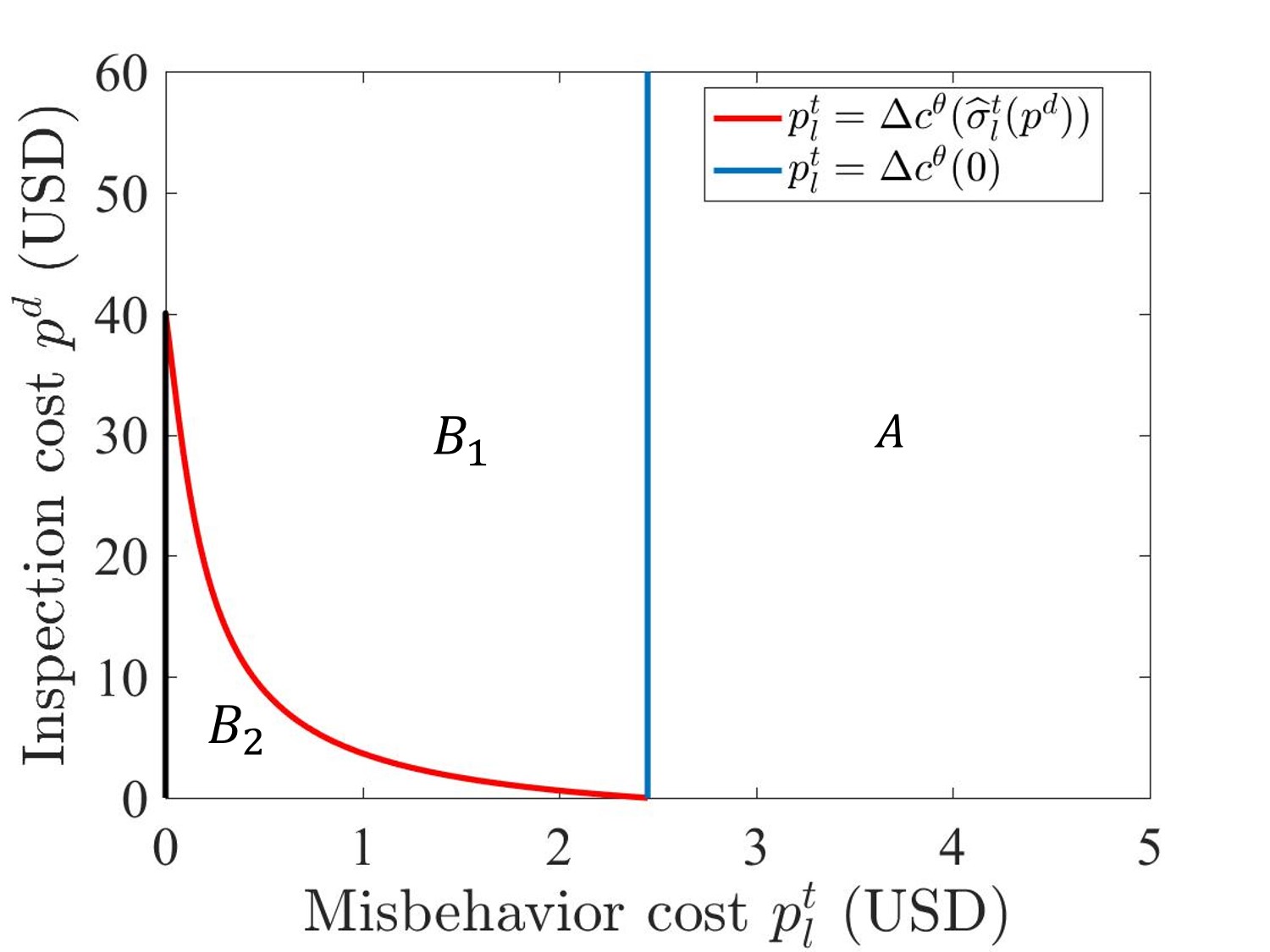}
    \caption{PBE regimes.}\label{fig:regime}
\end{figure}

We model the highway as two parallel $M/M/1$ queuing systems, one representing the tolled lanes ($\H$), and the other representing the toll-free lanes ($\L$). For background on modeling highway traffic with stochastic queuing models, see \cite{newell13}.
Both the tolled lanes and the toll-free lanes have a capacity (service rate) of 1700 veh/hr, i.e. $\mu_\H=\mu_\L=1700$ veh/hr.
The travel cost on a server is the product of the expected system time and the value of time $\mathrm{VoT}=50$ USD/hr. The fine is $\Fl=100$ USD. Following standard results in queuing theory, the expected cost functions are as follows:

\begin{footnotesize}
\begin{subequations}\label{queue}
\begin{align*}
&\ch(\sigma)=\left\{\begin{array}{ll}
\frac{\mathrm{VoT}}{\mu_\H-\theta\lambda(1-\sigh)-(1-\theta)\lambda\sigl}, &\quad  \text{if }\theta\lambda(1-\sigh)+(1-\theta)\lambda\sigl<\mu_\H,\\
\infty,&\quad  o.w.
\end{array}\right.\\
&\cl(\sigma)=\left\{\begin{array}{ll}
\frac{\mathrm{VoT}}{\mu_\L-(1-\theta)\lambda(1-\sigl)-\theta\lambda\sigh},  & \quad \text{if } (1-\theta)\lambda(1-\sigl)+\theta\lambda\sigh<\mu_\L,\\
\infty, & \quad o.w.
\end{array}\right.
\end{align*}
\end{subequations}
\end{footnotesize}
We can check that the cost functions satisfy Assumptions (A1) - (A3). Fig. \ref{fig:regime} illustrates the regimes of PBE.
\vspace{-0.3cm}
In this example, the minimum $\pl$ that deters misbehavior is $\dc(0)=2.35$ USD. Note that this is the technology cost per signal. A device that is used to manipulate the message sent to the RSU can be expensive, but if the device is repeatedly used, the average cost can be low.

Additionally, since the fine $\Fl=100>\dc(0)=2.35$, the sub-regime $\Bthree$ is empty. This implies that given any $\pl$ and $\pd$, the operator will not inspect all agents. Given parameters in $\Btwo$, $\pl=0.5$ USD and $\pd=5$ USD, the equilibrium misbehavior rate is $\siglwe=\siglbar=2.15\%$, and the inspection rate is $\sigdhwe=0.34 \%$.

\section{Concluding remarks}
In this article, we introduced a signaling game to study the effects of operator
inspection strategy on the strategic misbehavior in a V2I-based highway system. Our model captures three key features: (a) Travelers’ ability to exploit the vulnerabilities of V2I communications; (b) The incomplete and asymmetric information on the part of operator resulting from lack of direct observability of travelers' true type; (c) Congestion externality on the compliant travelers resulting from misbehavior. We provided a complete characterization of PBE, and derived comparative statics with respect to practically relevant parameters such as fine, costs of multi-priority lanes, and technological costs of inspection and misbehavior. These results suggest guidelines for the design and deployment of inspection technologies for smart transportation systems, and are relevant to a class of strategic integrity attacks, where deterrence via inspection is socially desirable.

\begin{appendices}
\section{Proof of Theorem \ref{theorem_equilibrium}}\label{proof_theorem}
\begin{enumerate}
\item[(a)] In regime $\A$, since $\pl >\dc(0)$, from \eqref{UlL}, we have $\EsigwelH \leq -\ch(\siglwe)-\pl < -\cl(\siglwe) =\EsigwelL$. Therefore, from \eqref{travel_low_deviate}, we must have $\siglwe=0$. From \eqref{bayes}, we can check that $\bwe(\h|\H)=1$ and $\bwe(\l|\H)=0$. Following from Lemma \ref{BR}, $\sigdhwe=0$. Thus, the PBE in \eqref{eq_A} is the unique equilibrium. 
\item[(b)] In regime $\B$, we first prove by contradiction that $\siglwe \in (0, 1)$. Assume that $\siglwe=0$, then from \eqref{eq_operator_rational} and \eqref{bayes}, $\bwe$ and $\sigdlwe$ must be in \eqref{eq_A}. Then, from \eqref{UlL}, $\EsigwelL=-\cl(0)$. However, if type $\l$ agents deviate to choose $\H$, the utility is $-\ch(0)-\pl$. Since in regime $\B$, $\pl <\dc(0)$, type $\l$ agents has incentive to deviate to $\H$, which contradicts $\siglwe=0$. Additionally, from Proposition \ref{no_pool}, $\siglwe<1$. Therefore, in this regimes $\siglwe \in (0,1)$, i.e. type $\l$ agents take both servers in equilibrium, which implies the follows:
\begin{align}\label{equal_l}
\EsigwelL=\EsigwelH.
\end{align}
Furthermore, there are three cases for $\sigdhwe$: $\sigdhwe=0$, $\sigdhwe \in (0, 1)$ and $\sigdhwe=1$. It turns out that these three cases correspond to subregime $\Bone$, $\Btwo$ and $\Bthree$, respectively. 

\begin{enumerate}
\item[($\Btwo$)] $\sigdhwe \in (0, 1)$:
In this case, from Lemma \ref{BR}, we know that $\bwe(\l|\H)$ must be $\pd/\Fl$, and  $\siglwe=\siglbar$ in \eqref{siglbar} is the unique equilibrium strategy. 
Additionally, from \eqref{equal_l}, the operator's strategy $\sigdlwe$ should satisfy $-\cl(\siglwe)=-\ch(\siglwe)-\pl-\Fl \sigdlwe$. Thus, $\sigdlwe$ is in \eqref{eq_B_2}. Furthermore, we can check that when $\pd$ and $\pl$ satisfy \eqref{B_2_condition}, the strategies and the beliefs in \eqref{eq_B_2_general} are all feasible. Therefore, PBE in \eqref{eq_B_2_general} is the unique equilibrium.

\item[($\Bone$)] $\sigdhwe=0$: In this case, From \eqref{equal_l}, we must have $\EsigwelH=-\ch(\siglwe)-\pl =-\cl(\siglwe)=\EsigwelL$, which leads to $\dc(\siglwe)=\pl$. From \eqref{bayes}, $\bwe$ is in \eqref{b_B_1}. 

We now argue that the strategy in \eqref{eq_B_1_general} is indeed a PBE if the cost parameters satisfy \eqref{B1}. We have argued that the condition $\pl <\dc(0)$ is needed to ensure that $\siglwe \in (0, 1)$. 

If $\pd>(1-\theta)\Fl$, then from Lemma \ref{BR}, we know that $\sigdhwe=0$. If $\pd<(1-\theta)\Fl$, then again from Lemma \ref{BR}, as long as $\siglwe < \siglbar$, $\sigdhwe=0$. Since $\dc(\sigl)$ decreases in $\sigl$ and $\dc(\siglwe)=\pl$, we must have $\pl =\dc(\siglwe)> \dc(\siglbar)$, which leads to constraints in \eqref{B1}.




\item[($\Bthree$)] $\sigdhwe =1$:
In this case, from \eqref{equal_l}, we obtain $\EsigwelL=-\cl(\siglwe)=-\ch(\siglwe)-\pl-\Fl=\EsigwelH$. Therefore, $\siglwe$ satisfies $\dc(\siglwe)=\pl+\Fl$. From \eqref{bayes}, $\bwe$ is obtained from \eqref{b_B_3}. To ensure that the action $\I$ is a dominant strategy for the operator, from Lemma \ref{BR}, we need $\pd<(1-\theta)\Fl$, and $\siglwe>\siglbar$. Besides, $\dc(\siglwe)$ decreases in $\siglwe$ and $\dc(\siglwe)=\pl+\F$, we can conclude that $\pl+\F=\dc(\siglwe)<\dc(\siglbar)$, i.e. $\pl$ satisfies \eqref{condition_B_3}. \QEDA
\end{enumerate} 
\end{enumerate}
\end{appendices}



\section*{Acknowledgments}
This work was supported in part by the Singapore National Research Foundation through the Singapore MIT Alliance for Research and Technology (SMART) Center for Future Mobility (FM), and US National Science Foundation (NSF) grants.

\bibliographystyle{ieeetr}
\bibliography{bib_LJ,bib_MXW}   

\begin{thebibliography}{10}

\bibitem{varaiya93}
P.~Varaiya, ``Smart cars on smart roads: Problems of control,'' {\em IEEE
  Transactions on Automatic Control}, vol.~38, no.~2, pp.~195--207, 1993.

\bibitem{papadimitratos09}
P.~Papadimitratos, A.~de~La~Fortelle, K.~Evenssen, R.~Brignolo, and S.~Cosenza,
  ``Vehicular communication systems: Enabling technologies, applications, and
  future outlook on intelligent transportation,'' {\em IEEE communications
  magazine}, vol.~47, no.~11, 2009.

\bibitem{kurzhanskiy15}
A.~A. Kurzhanskiy and P.~Varaiya, ``Traffic management: An outlook,'' {\em
  Economics of transportation}, vol.~4, no.~3, pp.~135--146, 2015.

\bibitem{bergenhem12}
C.~Bergenhem, S.~Shladover, E.~Coelingh, C.~Englund, and S.~Tsugawa, ``Overview
  of platooning systems,'' in {\em Proceedings of the 19th ITS World Congress,
  Oct 22-26, Vienna, Austria (2012)}, 2012.

\bibitem{jin18hscc}
L.~Jin, M.~\v{C}i\v{c}i\'c, S.~Amin, and K.~H. Johansson, ``Modeling impact of
  vehicle platooning on traffic: A fluid queueing approach,'' in {\em Hybrid
  Systems: Computation and Control, 21st ACM International Conference on}, ACM,
  2018.

\bibitem{petit15}
J.~Petit, F.~Schaub, M.~Feiri, and F.~Kargl, ``Pseudonym schemes in vehicular
  networks: A survey,'' {\em IEEE communications surveys \& tutorials},
  vol.~17, no.~1, pp.~228--255, 2015.

\bibitem{lawson08}
N.~Lawson, ``Highway to hell: Hacking toll systems,'' {\em Presentation at
  Blackhat}, 2008.

\bibitem{hubaux04}
J.-P. Hubaux, S.~Capkun, and J.~Luo, ``The security and privacy of smart
  vehicles,'' {\em IEEE Security \& Privacy}, vol.~2, no.~3, pp.~49--55, 2004.

\bibitem{alpcan2010network}
T.~Alpcan and T.~Ba{\c{s}}ar, {\em Network security: A decision and
  game-theoretic approach}.
\newblock Cambridge University Press, 2010.

\bibitem{spence1974market}
A.~M. Spence, {\em Market signaling: Informational transfer in hiring and
  related screening processes}, vol.~143.
\newblock Harvard University Press, 1974.

\bibitem{fudenberg1991perfect}
D.~Fudenberg and J.~Tirole, ``Perfect bayesian equilibrium and sequential
  equilibrium,'' {\em journal of Economic Theory}, vol.~53, no.~2,
  pp.~236--260, 1991.

\bibitem{newell13}
G.~F. Newell, {\em Applications of Queueing Theory}, vol.~4.
\newblock Springer Science \& Business Media, 2013.

\end{thebibliography}
\end{document}